\documentstyle[12pt]{article}
\newcommand{\be}{\begin{equation}}
\newcommand{\ee}{\end{equation}}
\newcommand{\ba}{\begin{eqnarray}}
\newcommand{\ea}{\end{eqnarray}}
\newcommand{\Z}{Z \!\!\! Z}
\newcommand{\sect}[1]{\begin{center}{\Large #1}\end{center}}
\newcommand{\ie}{{\it i.~e. }}
\newcommand{\for}{\;\;\; {\rm for} \;\;}
\newcommand{\Lie}{{\rm Lie} \,}
\newcommand{\z}{\partial_z}
\newcommand{\zb}{\partial_{\bar{z}}}
\newcommand{\vs}{\vspace*{3mm}}

\newcommand{\inttr}{\int d^2 z \, {\rm Tr}}
\newcommand{\Ao}{A^0_z}
\newcommand{\Aob}{A^0_{\bar{z}}}
\newcommand{\Ap}{A^\prime_z}
\newcommand{\Apb}{A^\prime_{\bar{z}}}
\newcommand{\hp}{\hat{\psi}_-}
\newcommand{\V}{V_z}
\newcommand{\Vb}{V_{\bar{z}}}
\newcommand{\th}{\tilde{h}}

\begin{document}

\pagestyle{empty} \noindent
\hspace*{115mm} \normalsize IASSNS-HEP-94/13\\ \hspace*{115mm} February 1994\\
\begin{center} \vspace*{30mm} \LARGE Mirror symmetry for the Kazama-Suzuki
models\\
\vspace*{20mm} \large M\aa ns Henningson\footnote[1]{Address after September 1,
1994: {\it Department of Physics, Yale University, New Haven, CT 06511}}\\
\vspace*{10mm} \normalsize \it School of Natural Sciences \\ Institute for
Advanced Study \\ Olden Lane \\ Princeton, NJ 08540 \\
\vspace*{40mm} \large Abstract \\ \end{center}
We study the $N = 2$ coset models in their formulation as supersymmetric gauged
Wess-Zumino-Witten models. A model based on the coset $G/H$ is invariant under
a symmetry group isomorphic to $\Z_{k+Q}$, where $k$ is the level of the model
and $Q$ is the dual Coxeter number of $G$. Using a duality-like relationship,
we show that the $\Z_m$ orbifold of the vectorially gauged model and the
$\Z_{\tilde{m}}$ orbifold of the axially gauged model are each others mirror
partners when $m \tilde{m} = k + Q$.

\newpage \pagestyle{plain} \sect{1. Introduction}

Mirror symmetry is an intriguing property of the space of $N = 2$ theories in
two dimensions \cite{Yau}. At the level of conformal field theory, it could be
formulated as an isomorphism between two theories, amounting to a change of
sign of the $U(1)$ generator which is part of, say, the left-moving $N = 2$
superconformal algebra. Trivial as this may seem, at the level of Lagrangian
representations of $N = 2$ theories it leads to quite remarkable conclusions.
In the case of sigma-models with Calabi-Yau target spaces, for example, mirror
symmetry states that topologically different target spaces (with opposite Euler
numbers) may give rise to equivalent physics.

Mirror symmetry has mostly been studied in the context of Calabi-Yau
sigma-models and Landau-Ginzburg effective field theories. This approach is
quite general, since it allows for the moduli of the theory to be varied.
However, in the Calabi-Yau case one is limited to theories of integer conformal
anomaly $\hat{c}$ which allow for a geometric interpretation, although mirror
symmetry seems to be more related to world-sheet than to space-time physics.
Furthermore, it has been difficult to understand the exact mirror map, since
the precise form of the Calabi-Yau and Landau-Ginzburg models is not known at
the critical point. In fact, the only rigorously established example of mirror
symmetry, the Greene-Plesser construction \cite{Greene-Plesser}, is based on
another type of model, namely tensor products of the exactly solvable $N = 2$
minimal models \cite{Cappelli-Itzykson-Zuber}. This example relies on certain
equivalences between orbifolds of the minimal models, which have been
established by algebraic methods \cite{Gepner-Qiu}\cite{Gepner}.

However, the minimal models could be regarded as a special case of the
Kazama-Suzuki models, \ie $N = 2$ coset models \cite{Kazama-Suzuki}. As such
they have a Lagrangian formulation in terms of supersymmetric gauged
Wess-Zumino-Witten models \cite{Schnitzer}. It should therefore be possible to
understand mirror symmetry for these models by a more conceptual path integral
argument. In fact, it has been conjectured that mirror symmetry for the minimal
models could be understood as a duality in the corresponding $SU(2)/U(1)$ coset
model \cite{Kiritsis}. Such duality transformations exchange chiral and twisted
chiral superfields, and therefore implement the change of sign of a $U(1)$
charge that characterizes mirror symmetry \cite{Rocek-Verlinde}.

The object of the present article is to make this conjectured relationship
precise. In section two, we review the Lagrangian formulation of the general
$G/H$ $N = 2$ coset models as supersymmetric gauged Wess-Zumino-Witten models.
In general, there are two inequivalent possibilities, corresponding to an axial
or vector gauging of a $U(1)$ factor. We show that these models are invariant
under a symmetry group isomorphic to $\Z_{k+Q}$, where $k$ is the level of the
model and $Q$ is the dual Coxeter number of $G$. In section three, we perform a
duality transformation which establishes mirror symmetry exactly to all orders
in $1/k$ between orbifolds with respect to subgroups of $\Z_{k+Q}$ of the
vectorially and axially gauged models.

\newpage \sect{2. The Lagrangian formulation of the Kazama-Suzuki models}

The $N=1$ coset models \cite{Goddard-Kent-Olive} constitute an important class
of superconformal field theories. Such a model is specified by a Lie group $G$
with a subgroup $H$ and a positive integer $k$ called the level. Kazama and
Suzuki \cite{Kazama-Suzuki} investigated the conditions under which these
models actually possess $N=2$ supersymmetry and found that this happens exactly
when the left coset space $G/H$ is a K\"ahler space. The most important
property for our purposes is that in these cases the group $H$ is always of the
form
\be
H \simeq H^0 \times H^\prime, \label{e1}
\ee
where $H^0$ is isomorphic to $U(1)$.

We will now give a Lagrangian description of these models as supersymmetric,
gauged Wess-Zumino-Witten models \cite{Schnitzer}. The $G$ Wess-Zumino-Witten
model \cite{Witten-1} is invariant under global (or even chiral) $G \times G$
transformations acting on the fundamental $G$ valued field $g$ as
\be
g \rightarrow ugv \for u, v \in G.
\ee
Not all of this symmetry may be gauged, however, but only so called
anomaly-free subgroups \cite{Witten-2}. We need to embed a subgroup $H$ of $G$
of the form (\ref{e1}) in $G \times G$ in an anomaly-free way. For the
$H^\prime$ factor of $H$ there is generically only one choice, namely
\be
g \rightarrow u^{-1} g u \for u \in H^\prime. \label{e21}
\ee
For the abelian factor $H^0$ of $H$ we have the two possibilities, usually
referred to as vector and axial gauging. We will treat these cases in parallel,
with the upper (lower) sign always referring to the vectorially (axially)
gauged model. The $H^0$ gauge transformations act as
\be
g \rightarrow u^{\mp 1} g u \for u \in H^0.
\ee

The supersymmetric gauged Wess-Zumino-Witten model contains bosonic fields $A$
and $g$ and fermionic fields $\psi_+$ and $\psi_-$. The field $A$ is a
connection on an $H$ bundle over the world-sheet. The fields $g$ and $\psi_+$
and $\psi_-$ are sections of associated bundles with typical fibers $G$ and
$\Lie G/H$, \ie the orthogonal complement of $\Lie H$ in $\Lie G$,
respectively. The gauge group acts on $g$ as we described in the previous
paragraph and on the fermionic fields as
\be
\psi_+ \rightarrow u^{-1} \psi_+ u \for u \in H
\ee
and
\ba
&& \psi_- \rightarrow u^{-1} \psi_- u \for u \in H^\prime \nonumber\\
&& \psi_- \rightarrow u^{\mp 1} \psi_- u^{\pm 1} \for u \in H^0.
\ea
Finally, the connection $A$ transforms in the usual way as
\be
A \rightarrow u^{-1} d u + u^{-1} A u \for u \in H, \label{e22}
\ee
where $d$ is the exterior derivative on the world-sheet.

The action of the model is
\be
S^\pm(g, A, \psi_+, \psi_-) = S^\pm_{\rm B}(g, A) + S^\pm_{\rm F}(A, \psi_+,
\psi_-), \label{e2}
\ee
where the bosonic part is given by
\ba
S^\pm_{\rm B}(g, A) & = & k S_{\rm WZW} (g) + \frac{k}{2 \pi} \inttr \left(
(\Apb + \Aob) g^{-1} \z g - (\Ap \pm \Ao) \zb g g^{-1} \right. \nonumber\\
&& \left. - \Ap \Apb - \Ao \Aob + (\Apb + \Aob) g^{-1} (\Ap \pm \Ao) g \right)
\label{e3}
\ea
and the fermionic part by
\ba
&& S^\pm_{\rm F}(A, \psi_+, \psi_-) \label{e4} \\
&& \hspace*{10mm} = \frac{ik}{4 \pi} \inttr \left( \psi_+ (\zb \psi_+ + [\Apb +
\Aob, \psi_+]) + \psi_- (\z \psi_- + [\Ap \pm \Ao, \psi_-]) \right). \nonumber
\ea
Here we have decomposed the connection as $A = A^0 + A^\prime$, where $A^0 \in
\Lie H^0$ and $A^\prime \in \Lie H^\prime$. The relevant properties of the
Wess-Zumino-Witten action $S_{\rm WZW}(g)$ are summarized by the
Polyakov-Wiegmann formula \cite{Polyakov-Wiegmann}
\be
S_{\rm WZW}(gh) = S_{\rm WZW}(g) + S_{\rm WZW}(h) - \frac{1}{2 \pi} \inttr
\left(g^{-1} \z g \zb h h^{-1} \right),
\ee
which is valid for $g, h \in G$. For the special case of $h \in H^0$ we have
\be
S_{\rm WZW}(h) = \frac{1}{2 \pi} \inttr \left(-\frac{1}{2} h^{-1} \z h \zb h
h^{-1} \right).
\ee

It is straightforward to verify that the action (\ref{e2}) is invariant at the
classical level under local $H$ gauge transformations acting as in
(\ref{e21})-(\ref{e22}), and we choose the measure in the fermionic path
integral so that this symmetry is non-anomalous. Furthermore, the model is
invariant under infinitesimal left- and right-moving $N = 1$ supersymmetry
transformations, with parameters $\epsilon_+$ and $\epsilon_-$ respectively,
acting as
\ba
\delta g & = & i \epsilon_- g \psi_+ + i \epsilon_+ \psi_- g \nonumber\\
\delta \psi_+ & = & \epsilon_- (1 - \Pi_H) (g^{-1} \z g + g^{-1} (\Ap \pm \Ao)
g - i \psi_+ \psi_+) \label{e5} \\
\delta \psi_- & = & \epsilon_+ (1 - \Pi_H) (\zb g g^{-1} - g(\Apb + \Aob)
g^{-1} + i \psi_- \psi_-) \nonumber\\
\delta A & = & 0. \nonumber
\ea
Here $\Pi_H$ is the orthogonal projection of $\Lie G$ on $\Lie H$. These
formulas are gauge covariant, although not manifestly so. By inspection of the
action (\ref{e2}), one may in fact show \cite{Witten-3}\cite{Nakatsu} that the
model is invariant under left- and right-moving $N = 2$ supersymmetry when
$G/H$ is a K\"ahler space.

Finally, we will consider transformations of the form
\ba
g & \rightarrow & \alpha g \nonumber\\
\psi_- & \rightarrow & \alpha \psi_- \alpha^{-1} \label{e6} \\
\psi_+ & \rightarrow & \psi_+ \nonumber\\
A & \rightarrow & A, \nonumber
\ea
where $\alpha \in H^0$ is a constant. These transformations commute with the
gauge transformations (\ref{e21})-(\ref{e22}) and the supersymmetry
transformations (\ref{e5}). We could have added an arbitrary gauge
transformation, but the form that we have given will allow us to carry out the
mirror symmetry transformation in the next section exactly to all orders in
$1/k$.

The action (\ref{e2}) is in general not invariant under the transformations
(\ref{e6}), though. The fermionic part (\ref{e4}) is invariant at the classical
level, but at the quantum level the symmetry breaks down due to the chiral
anomaly. The anomalous variation is
\be
S^\pm_{\rm F}(A, \psi_+, \psi_-) \rightarrow S^\pm_{\rm F}(A, \psi_+, \psi_-)
\mp \gamma \frac{Q}{2 \pi} \inttr \left(T^0 F^0_{z\bar{z}} \right),
\ee
where $F^0_{z\bar{z}} = \z \Aob - \zb \Ao$ is the $\Lie H^0$ component of the
curvature of $A$, and $Q$ denotes the dual Coxeter number of $G$. We have
defined $\gamma$ by $\alpha = \exp \gamma T^0$, where $T^0 \in \Lie H^0$ is
normalized so that $\exp 2 \pi T^0$ is the identity element of $H^0$.  Note
that although the action is proportional to $k$, the anomaly, being a one-loop
quantum effect, is independent of $k$.

It is not difficult to understand why $Q$ enters in the above expression: The
anomaly is proportional to the sum of the squares of the $U(1)$ charges of the
fermions. But the fermions, which take their values in $\Lie G/H$, transform
under $H^0$ in the restriction of the adjoint representation of $G$.
Furthermore, $\Lie H$ is uncharged under $H^0$ acting in this way. The anomaly
is therefore proportional to the sum of the squares of the charges of $G$ under
$H^0$ acting in the adjoint representation, \ie the dual Coxeter number of $G$.

We now turn to the bosonic part (\ref{e3}), which is non-invariant already at
the classical level and transforms under (\ref{e6}) as
\be
S^\pm_{\rm B}(g, A) \rightarrow S^\pm_{\rm B}(g, A) \mp \gamma \frac{k}{2 \pi}
\inttr \left(T^0 F^0_{z\bar{z}} \right).
\ee
There are no quantum corrections to this result, though, since we take the
integration measure for the path integral over $g$ to be a product of a Haar
measure for each point on the world-sheet, so that it is invariant under left
(and right) multiplication.

The complete action (\ref{e2}) thus transforms under (\ref{e6}) as
\be
S^\pm(g, A, \psi_+, \psi_-) \rightarrow S^\pm(g, A, \psi_+, \psi_-) \mp \gamma
(k + Q) c_1,
\ee
where $c_1 = \frac{1}{2 \pi} \inttr \left(T^0 F^0_{z\bar{z}} \right) \in \Z$ is
the first Chern class of the $U(1)$ bundle over the world-sheet on which $A^0$
is a connection. We see that the action changes by a multiple of $2 \pi$ if we
take $\gamma$ to be a multiple of $2 \pi / (k+Q)$. But this is as good as being
invariant, since the action only enters as $\exp i S$ in the path integral.
Note that the Wess-Zumino-Witten action $S_{\rm WZW}(g)$ is only well-defined
modulo a multiple of $2 pi$ in any case \cite{Witten-1}. Since $\gamma = 2 \pi$
corresponds to the identity transformation, we have thus shown that the model
is invariant under a $\Z_{k+Q}$ quantum symmetry.

It will be convenient to change variables according to
\ba
g & = & h t \\
\psi_- & = & h \hp h^{-1}, \nonumber
\ea
where $h \in H^0$. The decomposition of $g$ is clearly not unique, but amounts
to choosing a section $t$ of $G$ when $G$ is regarded as a $U(1)$ bundle over
the left coset space $G/H^0$. The transformations (\ref{e6}) now act as
\be
h \rightarrow \alpha h, \label{e7}
\ee
whereas the remaining fields $A$, $t$, $\psi_+$ and $\hp$ are invariant.
Because of the chiral anomaly, the change of variables from $\psi_-$ to $\hp$
gives rise to a Jacobian from the measure in the fermionic path integral. We
should therefore add the term
\be
S^\pm_{\rm c.a.}(h, A) = \frac{Q}{2 \pi} \inttr \left( -\frac{1}{2} h^{-1} \z h
\zb h h^{-1} \mp \Ao \zb h h^{-1} \right)
\ee
to the action (\ref{e2}). The complete action is thus
\ba
&& S^\pm_{\rm eff}(h, t, A, \psi_+, \hp) = S^\pm_{\rm B}(h t, A) + S^\pm_{\rm
F}(A, \psi_+, h \hp h^{-1}) + S^\pm_{\rm c.a.}(h, A) \nonumber\\
&& = k S_{\rm WZW}(t) + \frac{k}{2 \pi} \inttr \left( -\frac{1}{2} h^{-1} \z h
\zb h h^{-1} - h^{-1} \z h \zb t t^{-1} + (\Apb + \Aob) t^{-1} \z t \right.
\nonumber\\
&& \hspace*{50mm} + (\Apb + \Aob) t^{-1} h^{-1} \z h t - (\Ap \pm \Ao) \zb t
t^{-1} \mp \Ao \zb h h^{-1} \nonumber \\
&& \hspace*{50mm} \left. - \Ap \Apb - \Ao \Aob + (\Apb + \Aob) t^{-1} (\Ap \pm
\Ao) t \right) \label{e8} \\
&& \hspace*{3mm} + \frac{ik}{4 \pi} \inttr \left( \psi_+ (\zb \psi_+ + [\Apb +
\Aob, \psi_+]) + \hp (\z \hp + [\Ap \pm \Ao + h^{-1} \z h, \hp]) \right)
\nonumber\\
&& \hspace*{3mm} + \frac{Q}{2 \pi} \inttr \left( -\frac{1}{2} h^{-1} \z h \zb h
h^{-1} \mp \Ao \zb h h^{-1} \right). \nonumber
\ea

{}From a conformal field theory invariant under some discrete group of
symmetries, we may construct other conformal field theories by taking an
orbifold of the original theory \cite{Dixon-Harvey-Vafa-Witten}. In a
Lagrangian formulation, this amounts to allowing the fields to be well-defined
only modulo the action of the symmetry group. In our case, we may construct
orbifolds of the theory by allowing $h$ to be well-defined only modulo the
action of some subgroup of the $\Z_{k+Q}$ symmetry acting as in (\ref{e7}).

\newpage \sect{3. The mirror transformation}

In this section, we will show that the $\Z_m$ orbifold of the axially gauged
$G/H$ model is the mirror partner of the $\Z_{\tilde{m}}$ orbifold of the
vectorially gauged model when $m \tilde{m} = k + Q$, where $k$ is the level of
the model and $Q$ is the dual Coxeter number of $G$.

The proof of mirror symmetry is based on an argument which closely resembles
the familiar duality between non-linear sigma models with isometries
\cite{Rocek-Verlinde}. The isometry in our case corresponds to a global $U(1)$
transformation acting on the field $h$ as in (\ref{e7}). The crucial property
is that $h$ only enters in the combinations $\V = h^{-1} \z h$ and $\Vb = \zb h
h^{-1}$ in the action (\ref{e8}).
(However, as we have seen in the previous section, the action is not invariant
under a global $U(1)$ transformation acting on $h$, but rather transforms with
a term proportional to the first Chern class of the $U(1)$ bundle on which
$A^0$ is a connection.) We may therefore give an equivalent formulation of the
theory by changing variables from $h \in H^0$ to $\V$, $\Vb \in \Lie H^0$ in
(\ref{e8}) and adding the term
\be
S_{\tilde{\phi}}(\tilde{\phi}, V) = \frac{k+Q}{2 \pi} \inttr \left(
\frac{\tilde{\phi}}{2} T^0 (\z \Vb - \zb \V) \right).
\ee
Here $\tilde{\phi}$ is defined modulo $2 \pi / \tilde{m}$ and $T^0 \in \Lie
H^0$ is again normalized so that $\exp 2 \pi T^0$ is the identity element of
$H^0$. The resulting action is
\ba
&& S^\pm_1(\tilde{\phi}, V, t, A, \psi_+, \hp) = k S_{\rm WZW}(t) \label{e10}
\\
&& \hspace*{5mm} + \frac{k}{2 \pi} \inttr \left( -\frac{1}{2} \V \Vb - \V \zb t
t^{-1} + (\Apb + \Aob) t^{-1} \z t + (\Apb + \Aob) t^{-1} \V t \right.
\nonumber\\
&& \hspace*{10mm} \left. - (\Ap \pm \Ao) \zb t t^{-1} \mp \Ao \Vb - \Ap \Apb -
\Ao \Aob + (\Apb + \Aob) t^{-1} (\Ap \pm \Ao) t \right) \nonumber\\
&& \hspace*{5mm} + \frac{ik}{4 \pi} \inttr \left( \psi_+ (\zb \psi_+ + [\Apb +
\Aob, \psi_+]) + \hp (\z \hp + [\Ap \pm \Ao + \V, \hp]) \right) \nonumber\\
&& \hspace*{5mm} + \frac{Q}{2 \pi} \inttr \left( -\frac{1}{2} \V \Vb \mp \Ao
\Vb \right) + \frac{k+Q}{2 \pi} \inttr\left( \frac{\tilde{\phi}}{2} T^0 (\z \Vb
- \zb \V) \right). \nonumber
\ea
If we perform the path integral over $\tilde{\phi}$, we will get back the
original action (\ref{e8}). Namely, $\tilde{\phi}$ acts as a Lagrange
multiplier for the constraint $\z \Vb - \zb \V = 0$, which we may solve locally
as $\V = h^{-1} \z h$ and $\Vb = \zb h h^{-1}$ with $h \in H^0$. Although $V$
is flat it may have non-trivial holonomies around non-contractible cycles of
the world-sheet, \ie $h$ need not be well-defined globally. However, since
$\tilde{\phi}$ is a compact variable with period $2 \pi / \tilde{m}$, the
action (\ref{e10}) contains a term
\be
n_C m \oint_C {\rm Tr} \left( \frac{1}{2} T^0 V \right)
\ee
for each cycle $C$ of the world-sheet, where $n_C$ denotes the winding number
of $\tilde{\phi}$ around $C$ \cite{Rocek-Verlinde}. The path integral over
$\tilde{\phi}$ includes a sum over the winding numbers $n_C$, which will
constrain the holonomies $\oint_C V$ to be multiples of $2 \pi T^0 / m$. The
field $h \in H^0$ is thus well-defined modulo the action of the $\Z_m$ subgroup
of $U(1)$ acting as in (\ref{e7}), so the action (\ref{e10}) indeed describes
the $\Z_m$ orbifold of the original theory.

To obtain the mirror theory, we integrate out the field $V$ from the action
(\ref{e10}). The path integral over $\Vb$ gives a delta function with argument
\be
(k + Q) (-\V \mp 2 \Ao - T^0 \z \tilde{\phi}). \label{e9}
\ee
The path integral over $\V$ is now trivial to perform, and we get the dual
action
\ba
&& S^\pm_{\rm dual}(\th, t, A, \psi_+, \hp) \nonumber\\
&& = k S_{\rm WZW}(t) + \frac{k}{2 \pi} \inttr \left( -\frac{1}{2} \th^{-1} \z
\th \zb \th \th^{-1} - \th^{-1} \z \th \zb t t^{-1} + (\Apb + \Aob) t^{-1} \z t
\right. \nonumber\\
&& \hspace*{50mm} + (\Apb + \Aob) t^{-1} \th^{-1} \z \th t - (\Ap \pm \Ao) \zb
t t^{-1} \mp \Ao \zb \th \th^{-1} \nonumber \\
&& \hspace*{50mm} \left. - \Ap \Apb - \Ao \Aob + (\Apb + \Aob) t^{-1} (\Ap \pm
\Ao) t \right) \label{e11} \\
&& \hspace*{3mm} + \frac{ik}{4 \pi} \inttr \left( \psi_+ (\zb \psi_+ + [\Apb +
\Aob, \psi_+]) + \hp (\z \hp + [\Ap \pm \Ao + \th^{-1} \z \th, \hp]) \right)
\nonumber\\
&& \hspace*{3mm} + \frac{Q}{2 \pi} \inttr \left( -\frac{1}{2} \th^{-1} \z \th
\zb \th \th^{-1} \mp \Ao \zb \th \th^{-1} \right). \nonumber
\ea
Here we have introduced $\th = \exp - \tilde{\phi} T^0$, which thus is well
defined as an element of $H^0$ modulo the action of the $Z_{\tilde{m}}$
subgroup. We now note that
\be
S^\pm_{\rm dual}(\th, t, A, \psi_+, \hp) = S^\mp_{\rm eff}(\th, t, A, \psi_+,
\hp).
\ee

In many situations, such as the two-dimensional black hole solution to string
theory \cite{Witten-4} or the path integral derivation of duality for
non-linear sigma models \cite{Rocek-Verlinde}, the procedure of integrating out
non-dynamical vector fields is only correct classically. The reason is that the
coefficient of the $\V \Vb$ term in the analogue of (\ref{e10}) in general
depends on the other fields of the theory. This coefficient will appear in the
argument of the delta function in (\ref{e9}) and thus gives rise to a Jacobian
factor when the path integral over $\V$ is carried out.
To first order in $1/k$ this yields a non-trivial dilaton potential
\cite{Buscher}, but there will be further corrections to (\ref{e11}) to all
orders in $1/k$. In our case, however, the action (\ref{e11}) is correct to all
orders in $1/k$, since the coefficient of the $\V \Vb$ term in (\ref{e10}) is a
constant. This is the motivation for the particular choice of gauge that we
made in (\ref{e6}). We have thus shown that the $\Z_m$ orbifold of the axially
gauged model is the mirror partner of the $\Z_{\tilde{m}}$ orbifold of the
vectorially gauged model.

Finally, we remark that in the case of for example $G/H \simeq SU(2)/U(1)$ the
axially and vectorially gauged models are really equivalent, being related by a
change of variables which leaves the measure in the path integral invariant. We
thus recover mirror symmetry for the $N = 2$ minimal models
\cite{Gepner-Qiu}\cite{Gepner}. For a generic $G/H$, the axial and vector
gaugings are inequivalent, however.

\vs
I have benefitted from discussions with P. Berglund and E. Witten. This
research was supported by the Swedish Natural Science Research Council (NFR).

\end{document}